\documentclass[12pt,a4paper]{article}
\newcommand{\e}{\mathrm{e}}
\newcommand{\be}{\begin{equation}}
\newcommand{\ee}{\end{equation}}
\newcommand{\bea}{\begin{eqnarray}}
\newcommand{\eea}{\end{eqnarray}}

\begin{document}

\title{Lagrangian versus Quantization}

\author{Ciprian Acatrinei\thanks{On leave from: 
       {\it National Institute of Nuclear Physics and Engineering -
        P.O. Box MG-6, 76900 Bucharest, Romania}; e-mail:
        acatrine@physics.uoc.gr.} \\
        Department of Physics, University of Crete, \\
        P.O. Box 2208, Heraklion, Greece}

\date{December 27, 2002}

\maketitle

\begin{abstract}
We discuss examples of systems which can be quantized consistently,
although they do not admit a Lagrangian description. 
\end{abstract}

Whether a given set of equations of motion  admits or not a Lagrangian formulation
has been an interesting issue for a long time. 
As early as 1887, Helmholtz formulated necessary and sufficient conditions for this to happen, 
and the problem has a rich history \cite{invpb}. 
More recently, motivated by some unpublished work of Feynman \cite{Dyson},
a connection was made between the existence of a Lagrangian and the commutation relations 
satisfied by a given system  \cite{HS,Hughes}.
Ref. \cite{HS} concluded that under quite general conditions,
including commutativity of the coordinates, $[q_i,q_j]=0$, 
the equations of motion of a point particle admit a Lagrangian formulation.
The purpose of this note is to demonstrate the reverse, namely that 
noncommutativity of the coordinates forbids a Lagrangian formulation 
(therefore a Lagrangian implies commutativity). 
This happens in all but a few cases, which we all identify.
On the other hand, an extended Hamiltonian formulation always remains available.
It permits quantization of the system in any of the three usual formalisms: operatorial, wave-function, 
or path integral. Several examples will be used to illustrate the properties of such unusual systems.

We work in a (2+1)-dimensional space,
although our considerations easily extend to higher dimensions,
and  assume that
\be
[q_1,q_2]=i\theta\neq 0.
\ee
For generality, we allow for a nonzero commutator between the momenta,
$[p_1,p_2]=i\sigma$, 
in addition to the usual $[q_i,p_j]=i\delta_{ij}$ relations.
The commutation relations of interest are thus 
\be
\label{cr}
[x_a,x_b]=i \Theta_{ab}, \quad x_{1,2,3,4}=q_1,q_2,p_1,p_2 ,
\ee
with the constant antisymmetric matrix $\Theta_{ab}=(\omega^{-1})_{ab}$ given by
\be
\label{setup}
\Theta=
\left (
\begin{array}{rrrr}
0 & \theta & 1 & 0  \\
-\theta & 0 & 0 &1 \\
-1 &  0 & 0 & \sigma \\
0 & -1  & -\sigma &  0
\end{array}
\right )
\quad , \quad
\omega=\frac{1}{1-\theta\sigma}
\left (
\begin{array}{rrrr}
0 & \sigma &-1 & 0  \\
-\sigma & 0 & 0 &-1\\
1 &  0 & 0 &  \theta\\
0 & 1  &  -\theta & 0
\end{array}
\right ).  \label{omega}
\ee
We have denoted the phase space variables $q_1,q_2,p_1,p_2$ by $x_a$, $a=1,2,3,4$.
Eqs. (\ref{cr},\ref{setup}), together with a given Hamiltonian $H$,
completely determine the dynamics.

{\bf Classical dynamics: general}

At the classical level, Eqs. (\ref{cr},\ref{setup}) correspond to 
the following fundamental Poisson brackets 
\be
\{x_a,x_b\}=\Theta_{ab}.  \label{PB}
\ee
For two generic functions $A$ and $B$,
$\{A,B\}\equiv\frac{\partial A}{\partial x_a}\Theta_{ab}\frac{\partial B}{\partial x_b}$.
We will first show that  a dynamical system obeying (\ref{PB}) does not allow 
(in most cases) a Lagrangian formulation.

A classical system with Hamiltonian $H(x_i)$ and 
Poisson brackets (\ref{PB}) has the folowing equations of motion \cite{a1}
\be
\dot{x}_a=\{x_a,H\}=\Theta_{ab}\frac{\partial H}{\partial x_b}, \quad a,b=1,2,3,4.
\label{em}
\ee
More explicitely,
\be
\dot{q}_i=
\frac{\partial H}{\partial p_i}+\theta \epsilon_{ij}\frac{\partial H}{\partial q_j},
\qquad
\dot{p}_i=
-\frac{\partial H}{\partial q_i}+\sigma \epsilon_{ij}\frac{\partial H}{\partial p_j},
\qquad i,j=1,2.
\label{emq}
\ee
Above, $\epsilon_{12}=-\epsilon_{21}=1$. When $\theta=\sigma=0$, 
Eqs. (\ref{emq}) become the usual Hamilton equations.

We assume that $H=\frac{1}{2m}(p_1^2+p_2^2)+V(q_1,q_2)$.
(for kinetic terms of the form $(p_i-A_i(q))^2$, see \cite{a1}.)
The momenta are then given by
\be
p_i=m \dot{q}_i-m\theta \epsilon_{ij} \frac{\partial V}{\partial q_j}.
\label{momenta}
\ee 
Eliminating them from (\ref{emq}), one obtains the coordinate equations of motion,
\be
m\ddot{q}_i=-(1-\theta\sigma)\frac{\partial V}{\partial q_i}
+\sigma \epsilon_{ij} \dot{q}_j
+m \theta \epsilon_{ij} \frac{d}{dt}\frac{\partial V}{\partial q_j}, \quad i=1,2.
\label{em1}
\ee
As previously noted \cite{a1}, if $\theta\neq 0$, equations (\ref{em1})
are not in general derivable from a Lagrangian.
We will make this statement precise, through the use of the Helmholtz conditions. 
Those state \cite{invpb, Hughes, HS}
that a force $F_i$ is derivable from a Lagrangian, i.e.
$F_i=-\frac{\partial W}{\partial q_i}+\frac{d}{d t}\frac{\partial W}{\partial \dot{q}_i}$
where $W(q_i,\dot{q}_i,t)$,
if and only if $F_i$ is at most a linear function of the accelerations $\ddot{q}_i$,
and it satisfies:
\be
\frac{\partial F_i}{\partial \ddot{q}_j}=\frac{\partial F_j}{\partial \ddot{q}_i},
\qquad
\frac{\partial F_i}{\partial \dot{q}_j}+\frac{\partial F_j}{\partial \dot{q}_i}=
\frac{d}{dt} \left (
\frac{\partial F_i}{\partial \ddot{q}_j}+\frac{\partial F_j}{\partial \ddot{q}_i} \right ),
\label{Helmholtz}
\ee
\be
\frac{\partial F_i}{\partial q_j}-\frac{\partial F_j}{\partial q_i}=
\frac{1}{2}\frac{d}{dt} \left (
\frac{\partial F_i}{\partial \dot{q}_j}-\frac{\partial F_j}{\partial \dot{q}_i} \right ).
\ee
In our case 
the Helmholtz conditions reduce to
\be
\frac{\partial F_1}{\partial \dot{q}_2}+\frac{\partial F_2}{\partial \dot{q}_1}=0,
\qquad \frac{\partial F_1}{\partial \dot{q}_1}=\frac{\partial F_2}{\partial \dot{q}_2}=0,
\label{H1}
\ee
\be
\frac{\partial F_1}{\partial q_2}-\frac{\partial F_2}{\partial q_1}=
\frac{1}{2}\frac{d}{dt} 
\left (
\frac{\partial F_1}{\partial \dot{q}_2}-\frac{\partial F_2}{\partial \dot{q}_1}
\right ). \label{H2}
\ee
Eqs. (\ref{H1},\ref{H2})  constrain the potential $V$ in Eq. (\ref{em1}) to be of the form
\be
V(q_1,q_2,t)=\frac{1}{2}a(q_1^2+q_2^2)+b(t)q_1+c(t)q_2 \label{V}.
\ee
For generality, we allowed explicit time dependence of $V$.
This permits $b(t),c(t)$ to be arbitrary functions of time. 
The coefficient $a$ of the quadratic term is constrained by (\ref{H2}) to be constant.
Thus the most general equations of motion engendered by (\ref{cr}),
which do admit a Lagrangian description, are
\be
m\ddot{q}_1=-a(1-\theta\sigma)q_1 + (\sigma+\theta m a)\dot{q}_2+
[\theta m \dot{c}-(1-\theta\sigma)b], \label{l1}
\ee 
\be
m\ddot{q}_2=-a(1-\theta\sigma) q_2 - (\sigma+\theta m a)\dot{q}_1-
[\theta m \dot{b}+(1-\theta\sigma)c], \label{l2}
\ee
with $a$ constant and $b(t),c(t)$.
The right hand side term contains three types of solvable forces: 
harmonic oscillator, magnetic field, and homogeneous (possibly time-dependent).
The general solution of Eqs. (\ref{l1},\ref{l2}) can be found by standard methods.
We will discuss particular cases, which illustrate better their properties.
Of course, when $\theta=0$, $\sigma=0$, 
one gets the usual behaviour one expects from the potential (\ref{V}). 
Otherwise, some surprising effects appear.
First, even when $V=0$, one has an effective magnetic field $\sigma$
acting on the whole 2D plane. All the particles are equally charged under it.
Second, the external homogeneous force disappears not only if $b=c=0$,
but also if  
$ b=\beta\cos \gamma t, c=\beta \sin \gamma t$, and $\omega=(1-\theta\sigma)/\theta m$.
Thus, from a "commutative" point of view, one applies oscillatory forces along the directions 
$q_1$ and $q_2$, but no force is registerd due to noncommutativity (NC) of the coordinates!
Third, if $\sigma+\theta m a=0$, the magnetic-like force disappears.
Finally, if $1=\theta\sigma$, one has no Newton-like term at all. 
In this case the system undergoes a dimensional reduction. 
The system of differential equations (\ref{emq}) becomes degenerate 
and a first-order Lagrangian description exists \cite{dimred,a1}.

A few remarks are in order.
First, an interesting situation appears when $0<|1-\sigma\theta|<<1$,
and $\sqrt{\sigma}$ is  big enough with respect to the momentum scales appearing
in the potential $V$.   
Then, the dynamics in Eq. (\ref{em1}) is controled by the magnetic force  
$\epsilon_{ij}\sigma \dot{q}_j$, and the potential $V$ can be treated as a small perturbation.

Second, cf. Eqs. (\ref{momenta},\ref{em1},\ref{l1},\ref{l2}),
$\sigma$ and $\theta$ at least partially play the role of magnetic fields, 
in a way depending also on the potential $V$. 
"Primordial magnetic fields", which are of much interested nowadays, 
can thus be generated by simply assuming noncommutativity. 
Although those effective magnetic fields would be tiny,
they would be coherent over large distances, contributing to large scale (e.g. cosmological) dynamics. 

Third, a Lagrangiam formulation can still be constructed for noncommuting coordinates, at a certain price.
One can mix the $q$'s and $p$'s through linear noncanonical transformations which block-diagonalize
the symplectic form (\ref{omega}). This however transfers nonlinearity from the potential term to the kinetic term 
of the Hamiltonian, a highly undesirable feature. Another possibility \cite{deriglazov} is to double the number of 
degrees of freedom, write a first-order Lagrangian in the extended space, 
then get rid of the unphysical degrees of freedom via constrained quantization. 
The first-order Lagrangian looks however very muck like a Hamiltonian, and the constraint analysis
proceeds anyway in Hamiltonian form.

{\bf Classical dynamics: examples}

We proceed with examples which do not admit a Lagrangian formulation,
and display some of their features.

Consider first the anisotropic harmonic oscillator potential, 
$V=\frac{1}{2}(a_1q_1^2+a_2q_2^2)$, which gives the equations of motion 
\be
m\ddot{q}_1=-(1-\theta\sigma)a_1q_1+(\sigma+\theta  m a_2)\dot{q}_2, \label{o1}
\ee
\be
m\ddot{q}_2=-(1-\theta\sigma)a_2q_2-(\sigma+\theta  m a_1)\dot{q}_1.\label{o2}
\ee
If we chose $\sigma+m\theta a_2=0$, then $\sigma+m\theta a_1\neq 0$, provided $a_1\neq a_2$.
$q_1$ becomes a harmonic oscillator,
whereas $q_2$ is a harmonic oscillator driven by a periodic force
$m\theta (a_1-a_2) \dot{q}_1$.
The solution for $q_1$ is the usual one,
$q_1(t)=q_1(0)\cos \omega_1 t +(q'_1(0)/\omega_1) \sin \omega_1 t  $,
whereas for $q_2$ it reads 
\be
q_2(t)=q_2(0)\cos \omega_2 t +\frac{q'_2(0)}{\omega_2} \sin \omega_2 t 
+\theta m\frac{q'_1(0) \cos \omega_1 t - \omega_1 q_1(0)\sin \omega_1 t}{1-\theta\sigma}.
\label{pert}
\ee
Above, $m\omega^2_i=(1-\theta\sigma)a_i$, $i=1,2$.
If $\theta$ is small, the last term in Eq.(\ref{pert}) is a perturbation
which produces oscillations around the commutative trajectory. 
The particle goes on a wiggly path, which averages to the commutative one.
If $\theta$ is big, or if $|1-\theta\sigma|<<1$, the "perturbation" explodes and dominates
the dynamics, which becomes completely different from the commutative one.
One sees a qualitative difference between a NC isotropic oscillator (which admits a Lagrangian form) 
and a NC anisotropic one (no Lagrangian form).

As a second example consider, commutatively speaking,
a constant force along $q_2$, and a harmonic one along $q_1$,
$V=\frac{1}{2}a_1 q_1^2+bq_2$.
The equations of motion are
\be
m\ddot{q}_1=-(1-\theta\sigma)a_1q_1+\sigma\dot{q}_2,  \label{f1}
\ee
\be
m\ddot{q}_2=-(1-\theta\sigma)b-(\sigma+\theta  m a_1)\dot{q}_1. \label{f2}
\ee
If $\sigma =0$, again $q_1$ is a harmonic oscillator, while $q_2$ is driven
by a constant plus periodic force.
The solution is the usual harmonic oscillator for $q_1$, while for $q_2$ one has
$$
q_2(t)=q_2(0)+[q'_2(0)+q_1(0)\theta a_1]t- \frac{bt^2}{2m}-
$$
\be
-\theta a_1\left[ \frac{q_1(0)}{\omega_1} \sin \omega_1 t
-\frac{q'_1(0)}{\omega_1^2} (1-\cos \omega_1 t)\right ].
\ee 
Again, the NC trajectory wiggles around the commutative one.
On the other hand, if $\sigma+\theta  m a_1=0$,
$q_2$ feels a constant force,  while the oscillator $q_1$ is driven
by a linearly time-dependent force $\sigma\dot{q}_2$.
One has the solution $q_2(t)=q_2(0)+t q'_2(0)-(1-\theta\sigma)\frac{bt^2}{2m}$,
but 
\be
q_1(t)=q_1(0)\cos \omega_1 t +\frac{q'_1(0)}{\omega_1} \sin \omega_1 t   
+ \frac{\sigma}{a_1}\left [\frac{q'_2(0)}{(1-\theta\sigma)}-\frac{b}{m}t \right ] \label{f3}
\ee
A drastic change occurs: $q_1$ grows linearly with time (it is not bounded anymore), 
and oscillates around this path as a commutative oscillator.

As a third example, consider a potential which depends only on one coordinate,
say $V=V(q_1)$. If $\sigma =0$ the equations of motion are
\be
m\ddot{q}_1=-\partial_1 V, \qquad  m\ddot{q}_2=-\theta m \frac{d}{dt}\partial_1 V=-\theta m^2 
\frac{d^3q_1}{dt^3}. \label{V1}
\ee
If $\theta\neq 0$, $q_1$ transfers nontrivial dynamics to $q_2$.
More precisely, once $q_1(t)$ is known
(its implicit form is $t(q_1) =\int_{0}^{q_1} \frac{dq'}{\sqrt{V(0)-V(q')}}$),
$q_2$ is fixed by the second equation in (\ref{V1}).
To illustrate, consider the quartic potential $V(q_1)=V(0)-\frac12 m^2 q_1^2+g q_1^4$.
One can not find simple expressions for $q_1(t)$ in a nonlinear problem in general.
However, the classical solution satisfying $q_1(t=-\infty)=0$ and
$q_1(t=0)=\frac{m}{\sqrt{g}}=\lambda$ is simple enough
\begin{equation}\label{cl_sol}
q_1(t)=\frac{m}{\sqrt{g}}\frac{2\e^{-mt}}{1+\e^{-2mt}}.
\end{equation}
Calculating $q_2(t)$ via (\ref{V1}) one obtains
\be
q_2(t)=q_2(0)+q'_2(0) t -\theta m \dot{q}_1(t),
\ee
radically different from the $\theta =0$ expression, $q_2(t)=q_2(0)+q'_2(0) t$.

Time-dependent backgrounds appearing "out-of-nowhere" are thus possible
in NC dynamics, see also Eqs. (\ref{l1},\ref{l2}). 

{\bf Quantization: formalism}

We have shown that, except for isotropic quadratic terms 
and linear couplings (constant forces), no Lagrangian formulation is available on NC spaces.
We discuss now the quantization of such systems.

Operatorial quantization is trivially implemented using
Eqs (\ref{cr},\ref{setup}):
\be
\frac{d}{dt}\hat{x}_a=i[\hat{x}_a,H]=i[\hat{x}_a,\hat{x}_b]\frac{\partial H}{\partial \hat{x}_b}
=\Theta_{ab}\frac{\partial H}{\partial \hat{x}_b}. \label{heisenberg}
\ee
The equations of motion (\ref{heisenberg}) are an extension of the usual Heisenberg ones.
They are the same as (\ref{em}), with the coordinates becoming operators. 

A phase space path integral for systems obeying the commutation relations (\ref{cr}) 
was constructed in  \cite{a2}. We do not repeat it here.

A Schr\"{o}dinger (wave function) formulation can be constructed 
as follows. First, chose a basis in the Hilbert space on which
the operators $\hat{x}_a$ act, for instance $|q_1,p_2>$, i.e. the eigenstates
of the operators $\hat{q}_1$ and $\hat{p}_2$.
Second, for an arbitrary state $|\psi>$, define the wave function
(half in coordinate space, half in momentum space)
\be
\psi(q_1,p_2,t)\equiv <\psi(t)|q_1,p_2>.
\ee
The commutation relations (\ref{cr})
imply that the operators $\hat{q}_2$ and $\hat{p}_1$ have the following action on $\psi$:
\be
\hat{q}_2\psi=i(\partial_{p_2}-\theta\partial_{q_1})\psi,
\qquad
\hat{p}_1\psi=i(-\partial_{q_1}+\sigma \partial_{p_2})\psi.
\label{repr1}
\ee
If $H=\frac{1}{2m}(\hat{p}_1^2+\hat{p}_2^2)+V(\hat{q}_1,\hat{q}_2)$, 
(\ref{repr1}) leads to the Schr\"{o}dinger equation
\be
i\frac{d}{dt}\psi=H\psi=
\left [
\frac{1}{2m}
\left (
p_2^2-(\partial_{q_1}-\sigma\partial_{p_2})^2
\right )
+V(q_1,i\partial_{p_2}-i\theta\partial_{q_1})
\right ]\psi(q_1,p_2).
\ee
If $\sigma =0$, a momentum-space wave function $\psi(p_1,p_2,t)$ also exists;
it will be discussed later.

{\bf Quantization: examples}

For an harmonic potential, it can be shown by path integrals \cite{a2}, or operatorially \cite{np}, 
that the only change induced by NC is  an anisotropy of the oscillator.
However, {\it starting} with an anisotropic oscillator,
$V=\frac{1}{2}(a_1q_1^2+a_2q_2^2)$, $a_1 \neq a_2$, makes an important difference. 
The  equations of motion are the same as in (\ref{o1},\ref{o2}), with  $q_{1,2}$ operators.
For simplicity, assume $\sigma+m\theta a_2=0$; then $\sigma+m\theta a_1\neq 0$.
$\hat{q}_2$ is driven by a periodic force and, being of the form (\ref{pert}),
transitions between the states of the quantum system will appear.

Our second example, $V=\frac{1}{2}a_1 q_1^2+bq_2$, also exhibits peculiar behaviour.
If $\sigma =0$, the operator solutions of (\ref{f1},\ref{f2}) 
again involve transitions which would be absent if $\theta =0$.
If $\sigma+\theta  m a_1=0$, changes are more dramatic. Eq. (\ref{f3}) shows 
that the particle  is not bounded anymore along $q_1$, in contrast with the commutative case.

Third, consider the case in which the potential depends only on one coordinate,
$V=V(q_1)$. If $\sigma =0$ an interesting phenomenon takes place. 
The commutation relations (\ref{cr}) admit a representation
in the basis $|p_1,p_2>$, $\psi(p_1,p_2,t)\equiv <\psi(t)|p_1,p_2>$:
\be
\hat{q_1}\psi=(i\partial_{p_1}+\theta \alpha p_2)\psi,
\qquad
\hat{q_2}\psi=(i\partial_{p_2}+\theta (1+\alpha) p_1)\psi(p_1,p_2),
\label{repr2}
\ee
with $\alpha$ a parameter,
and the Schr\"{o}dinger equation becomes
\be
i\frac{d}{dt}\psi=
\left [
\frac{1}{2m}
\left (p_1^2+p_2^2\right )
+V(i\partial_{p_1}+\theta \Lambda p_2,i\partial_{p_2}+\theta (1+\Lambda) p_1)
\right ]\psi(p_1,p_2)
\ee
This equation is (gauge) invariant under shifts of $\alpha$ by $\Lambda $,
\be
\alpha\rightarrow \alpha-\Lambda
\ee 
combined with multiplications 
of the momentum-space wave-function by a phase $e^{i\Lambda\theta p_1 p_2}$,
\be
\psi(p_1,p_2)\rightarrow e^{i\Lambda\theta p_1 p_2}\psi(p_1,p_2).
\ee
$\theta$ plays the role of a "magnetic field" in momentum space.

In particular, when $\Lambda=\alpha$, $\hat{q}_1$ becomes $\theta$-independent.
Then, if $V=V(q_1)$, the Schr\"{o}dinger equation is $\theta$-independent.
It has consequently the same spectrum with the commutative problem,
although classically the NC system does not even admit a Lagrangian formulation!
For example, $V(q_1,q_2)=V(q_1)=V(0)-\frac12 m^2 q_1^2+g q_1^4$, on a NC space,
gives rise to a nonlinear system without classical Lagrangian formulation, cf. (\ref{V}),
but which has the same spectrum as the corresponding commutative (Lagrangian) system.
 
If $V=V(q_1,q_2)$ the above gauge invariance persists, but does not eliminate $\theta$ from the wave equation.

\vskip 0.3cm

We conclude (in opposition with the spirit of \cite{HS})
that non-Lagrangian systems can be consistently quantized.
The formalism truly relevant for their quantization is the Hamiltonian one.
The examples we used to illustrate this point appear to have an interesting, 
or at least intriguing, behaviour.

\vskip 0.3cm

{\bf Acknowledgements}

I thank Kostas Anagnostopoulos and Theodore Tomaras for useful discussions.
This work was supported through a European Community Marie Curie fellowship,
under Contract HPMF-CT-2000-1060.


\end{document}